\documentstyle[12pt]{article}
\begin{document}\thispagestyle{empty}\begin{flushright}
OUT--4102--66\\MZ--TH/96--37\\hep-th/9612011\\
December  1997
            \end{flushright} \vspace*{2mm} \begin{center} {\Large\bf
Feynman diagrams as a weight system:\\[5pt]
four-loop test of a four-term relation$^{*)}$}\vglue 10mm{\large{\bf
D.~J.~Broadhurst$            ^{1)}$}\vglue 4mm
Physics Department, Open University\\
Milton Keynes MK7 6AA, UK           \vglue 6mm{\bf
D.~Kreimer$                ^{2,3)}$}\vglue 4mm
Institut f\"{u}r Physik, Johannes Gutenberg-Universit\"{a}t\\
Postfach 3980, D-55099 Mainz, Germany}\end{center}\vfill{\bf Abstract}\quad
At four loops there first occurs a test of the four-term
relation derived by the second author in the course of investigating
whether counterterms from subdivergence-free diagrams form a weight
system. This test relates counterterms in a four-dimensional field
theory with Yukawa and $\phi^4$ interactions, where no such
relation was previously suspected. Using integration by parts, we reduce
each counterterm to massless two-loop two-point integrals.
The four-term relation is verified, with
$\langle G_1-G_2+G_3-G_4\rangle=0-3\zeta_3+6\zeta_3-3\zeta_3=0$,
demonstrating non-trivial cancellation of the trefoil knot
and thus supporting the emerging connection between knots
and counterterms, via transcendental numbers assigned
by four-dimensional field theories to chord diagrams.
Restrictions to scalar couplings and renormalizable interactions are found
to be necessary for the existence of a pure four-term relation.
Strong indications of richer structure are given at five loops.
                                                    \vfill
                                                    \footnoterule\noindent
$^*$) work supported in part by grant CHRX--CT94--0579, from HUCAM\\
$^1$) email: D.Broadhurst@open.ac.uk\\
$^2$) email: kreimer@higgs.physik.uni-mainz.de\\
$^3$) Heisenberg Fellow
\newpage
\setcounter{page}{1}
\newcommand{\df}[2]{\mbox{$\frac{#1}{#2}$}}
\newcommand{\Eq}[1]{~(\ref{#1})}
\newcommand{\Eqq}[2]{~(\ref{#1},\ref{#2})}
\newcommand{\Eqqq}[3]{~(\ref{#1},\ref{#2},\ref{#3})}
\newcommand{\ct}[1]{\langle G_{#1}\rangle}
\newcommand{\ctt}[1]{\langle\tilde{G}_{#1}\rangle}
\newcommand{\ctb}[1]{\langle\overline{G}_{#1}\rangle}
\newcommand{\ep}{\varepsilon}
\newcommand{\psla}[1]{p\llap{/\kern-0.5pt}_{#1}}
\setlength{\unitlength}{0.014cm}
\newbox\shell
\newcommand{\blob}{\circle*{10}}
\newcommand{\lbl}[3]{\put(#1,#2){\makebox(0,0)[b]{$#3$}}}
\newcommand{\dia}[1]{\setbox\shell=\hbox{\begin{picture}(250,280)(-125,-140)#1
\end{picture}}\dimen0=\ht\shell\multiply\dimen0by7\divide%
\dimen0by16\raise-\dimen0\box\shell}

\subsection*{1. Introduction}

In~\cite{4TR} one of us (DK) formulated an argument leading to
the conclusion that a four-term relation is obeyed by
a class of subdivergence-free
counterterms obtainable by conventional perturbative expansions
of {\em bona fide\/} field theories, thus extending consideration
of four-term relations from the rarefied realm of topological~\cite{BN} field
theory to the concrete workbench of calculational~\cite{CT} techniques,
of practical value in four-dimensional spacetime.

There are two avenues opened up by the argument of~\cite{4TR}.
The first concerns the mapping~\cite{DK1,DK2} from knots to numbers,
realized~\cite{BKP,EUL,BGK,BK15} by counterterms. We remark that the discovery
of a four-term relation offers a prospect of deriving a knot-to-number
connection from the abstract properties of the resulting weight system.
It may thus provide {\em post hoc\/} clarification of the field-theoretic
successes~\cite{BKP,EUL,BGK,BK15,BDK,BKa,DKT,JJN,DK3,DEM} of the
ideas in~\cite{DK1,DK2}.
The second, here addressed, concerns tests of the four-term
relation and investigation of whether it fails when the stipulations
in~\cite{4TR} are not met.

In Section~2 we prosecute a successful test in a combined Yukawa
and $\phi^4$ theory, at four loops. Sections~3 and~4
confirm the expectations~\cite{4TR} that
a pure four-term relation is vitiated by vector couplings,
and by non-renormalizable interactions.
Section~4 also considers a specific three-term relation,
derived in~\cite{4TR}.
Section~5 offers conclusions.

\subsection*{2. Four terms, four loops, and four dimensions}

Fig.~1 shows the four subgraphs that generate every four-term relation.
In each of the four cases, three arcs of a circle are indicated,
with a chord connecting the upper pair.
These arcs form part of a hamiltonian circuit that passes through
every vertex of each diagram.
The connections of vertices on other parts of the hamiltonian circuit
need not yet concern us.
{}From the bottom arc, connections are made, in turn, to the four parts of
the hamiltonian circuit that are adjacent to the chord.
We assume that the four terms:
\begin{enumerate}
\item[(i)] are free of subdivergences;
\item[(ii)] differ only by the subgraphs of Fig.~1;
\item[(iii)] have trivial vertices, involving no vectorial (or
higher tensorial) structure;
\item[(iv)] involve no propagator with spin $s>\frac12$;
\item[(v)] modify one of the dimensionless couplings
of a renormalizable theory.
\end{enumerate}
\begin{center}{\bf Fig.~1}\quad
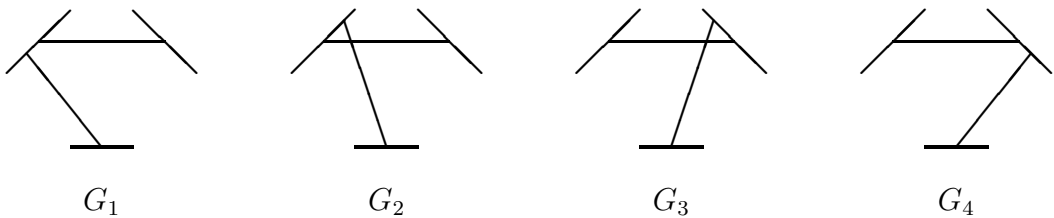
Every four-term relation contains these subgraphs.
\end{center}\vspace*{-5mm}\noindent\hspace*{5mm}
\dia{\thicklines\put(30,80){\line(1,-1){60}}\put(-90,20){\line(1,1){60}}
\put(-60,50){\line(1,0){120}}\put(-30,-50){\line(1,0){60}}\put(0,-50)
{\line(-4,5){71}}\lbl{0}{-110}{G_1}}\hfill
\dia{\thicklines\put(30,80){\line(1,-1){60}}\put(-90,20){\line(1,1){60}}
\put(-60,50){\line(1,0){120}}\put(-30,-50){\line(1,0){60}}\put(0,-50)
{\line(-1,3){40}}\lbl{0}{-110}{G_2}}\hfill
\dia{\thicklines\put(30,80){\line(1,-1){60}}\put(-90,20){\line(1,1){60}}
\put(-60,50){\line(1,0){120}}\put(-30,-50){\line(1,0){60}}\put(0,-50)
{\line(1,3){40}}\lbl{0}{-110}{G_3}}\hfill
\dia{\thicklines\put(30,80){\line(1,-1){60}}\put(-90,20){\line(1,1){60}}
\put(-60,50){\line(1,0){120}}\put(-30,-50){\line(1,0){60}}\put(0,-50)
{\line(4,5){71}}\lbl{0}{-110}{G_4}}\hspace*{5mm}\newpage\noindent
The necessity of this set of provisos is not established. In~\cite{4TR}
it is, however, claimed to be {\em sufficient\/} to derive the four-term
relation
\begin{equation}
\langle G_1-G_2+G_3-G_4 \rangle=0\label{4tr}
\end{equation}
where $\ct{k}$ is the corresponding counterterm,
i.e.\ the coefficient of overall logarithmic divergence
of the $k$-th of the four diagrams, numbered in cyclic order,
as in Fig.~1. These counterterms may be calculated by nullifying external
momenta and internal masses, and cutting the diagram wheresoever one pleases,
since infrared problems are excluded by the provisos. Thus if we can find a
non-trivial case, with less than five loops, in four dimensions, the machinery
of~\cite{CT} suffices to test the prediction\Eq{4tr}, without any subtleties
of infrared rearrangement.

The four-term relation of Fig.~1 necessarily operates on counterterms
with at least three loops, since it entails a hamiltonian circuit,
a chord, and a connection from an origin to one of the four
parts of the hamiltonian circuit that are adjacent to that chord.
To prevent subdivergences in four dimensions, there must be
at least one further loop. Indeed we have found only one four-loop
four-dimensional case in which the above conditions are satisfied.
There are several five-loop cases, but their computation
lies beyond what is systematically achievable by the algorithms
of~\cite{CT}.

\begin{center}{\bf Fig.~2}\quad
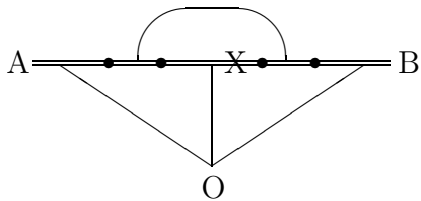
To generate four terms, at four loops, connect O to each blob, in turn.
\end{center}\vfill\begin{center}
\begin{picture}(400,200)(-200,-50)\put(0,100){\oval(140,100)[t]}
\put(-170,100){\line(1,0){340}}\put(-170,96){\line(1,0){340}}
\put(0,0){\line(0,1){96}}\put(0,0){\line(3,2){144}}\put(0,0){\line(-3,2){144}}
\put(-48,98){\blob}\put(-98,98){\blob}\put(48,98){\blob}\put(98,98){\blob}
\put(-195,88){A}\put(177,88){B}\put(-10,-30){O}\put(12,88){X}\end{picture}
\end{center}\vfill

To generate the four-loop test, consider Fig.~2, whose
four blobs indicate the connections that will be made
to the origin O.
The horizontal double line represents the propagation of a Dirac
fermion field,
$\psi$, with a Yukawa coupling, $\overline{\psi}\phi\psi$, to
a scalar boson field, $\phi$. At X there is a Yukawa coupling to an
external boson, which prevents subdivergences.
The asymmetry which it introduces also
guarantees non-triviality of the four-term relation.
Now we connect the origin O to each of the four blobs,
in turn, so that O becomes a $\phi^4$ vertex. Masses are then
set to zero, and the external momenta at A, B and X are nullified,
to give the four terms of Fig.~3.
Each has a (possible) overall logarithmic divergence, since it is a
contribution to the renormalization of the Yukawa coupling, which,
like the $\phi^4$ coupling, is dimensionless. After nullification,
we cut the four diagrams at convenient places, marked by $|$ in Fig.~3.
The value of each counterterm is thus given by a finite three-loop
massless two-point function. Moreover, the counterterms $\ct{1}$ and $\ct{4}$
factorize into products of one-loop
and two-loop functions. Hence we obtain $\ct{1,4}$ as two-loop integrals
and $\ct{2,3}$ as three-loop integrals. The latter may be reduced
to two-loop integrals, using integration by parts~\cite{CT}.

\newpage\begin{center}{\bf Fig.~3}\quad
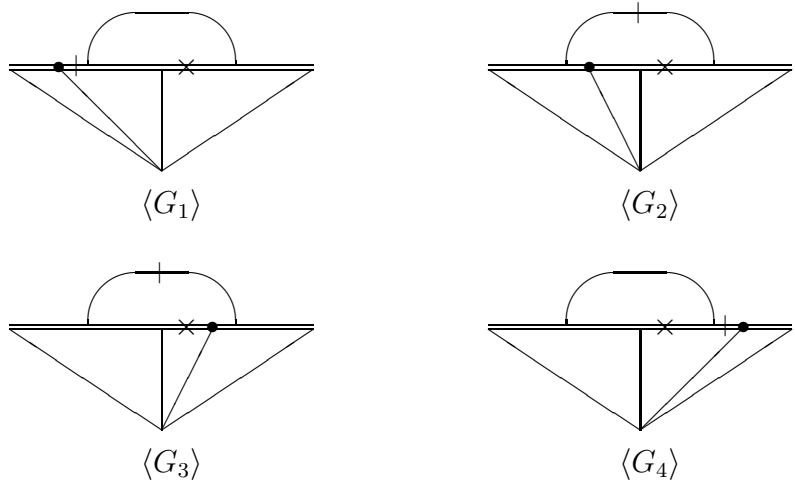
The four terms, after nullification, with cuts at convenient places.
\end{center}\noindent\vspace{20mm}\hspace*{3cm}
\dia{\put(-144,100){\line(1,0){288}}\put(-144,96){\line(1,0){288}}\put(0,100)
{\oval(140,100)[t]}\put(0,0){\line(0,1){96}}\put(0,0){\line(3,2){144}}
\put(0,0){\line(-1,1){96}}\put(0,0){\line(-3,2){144}}\put(-84,95)
{${\scriptstyle|}$}\put(-98,98){\blob}\put(-20,-40){$\ct{1}$}\put(12,91)
{$\times$}}\hfill
\dia{\put(-144,100){\line(1,0){288}}\put(-144,96){\line(1,0){288}}\put(0,100)
{\oval(140,100)[t]}\put(0,0){\line(0,1){96}}\put(0,0){\line(3,2){144}}
\put(0,0){\line(-1,2){48}}\put(0,0){\line(-3,2){144}}\put(-5,145)
{${\scriptstyle|}$}\put(-48,98){\blob}\put(-20,-40){$\ct{2}$}\put(12,91)
{$\times$}}\hspace*{3cm}\vspace*{-25mm}\\\hspace*{3cm}
\dia{\put(-144,100){\line(1,0){288}}\put(-144,96){\line(1,0){288}}\put(0,100)
{\oval(140,100)[t]}\put(0,0){\line(0,1){96}}\put(0,0){\line(1,2){48}}
\put(0,0){\line(3,2){144}}\put(0,0){\line(-3,2){144}}\put(-5,145)
{${\scriptstyle|}$}\put(48,98){\blob}\put(-20,-40){$\ct{3}$}\put(12,91)
{$\times$}}\hfill
\dia{\put(-144,100){\line(1,0){288}}\put(-144,96){\line(1,0){288}}\put(0,100)
{\oval(140,100)[t]}\put(0,0){\line(0,1){96}}\put(0,0){\line(1,1){96}}
\put(0,0){\line(3,2){144}}\put(0,0){\line(-3,2){144}}\put(78,95)
{${\scriptstyle|}$}\put(98,98){\blob}\put(-20,-40){$\ct{4}$}\put(12,91)
{$\times$}}\hspace*{3cm}\par\vspace*{-5mm}

Explicit expressions for the four counterterms may be compactly
written using
\begin{equation}
d\mu_n=\frac{(p_0^2)^{1+n\ep}}{(p_n-p_0)^2}
\prod_{k=1}^n\frac{d^{D}p_k}{\pi^{D/2}}
\,\frac{G(1+\ep)}{\left[G(1)\right]^2}\,
\frac{1}{p_k^2}\,\frac{1}{(p_{k-1}-p_k)^2}\label{mu}
\end{equation}
as a $n$-loop integration measure in $D\equiv4-2\ep$ euclidean dimensions,
with $p_0$ as the cut momentum, and
$G(\alpha)\equiv\Gamma(D/2-\alpha)/\Gamma(\alpha)$.
The four terms of Fig.~3 are given by
\begin{eqnarray}
\ct{1}&=&\df14\lim_{\ep\to0}{\rm Tr}\int d\mu_2
\frac{1}{\psla1}\psla{02}\label{G1}\\
\ct{2}&=&\df14\lim_{\ep\to0}{\rm Tr}\int d\mu_3
\frac{1}{\psla{10}}\psla1\psla2\frac{1}{\psla{30}}\label{G2}\\
\ct{3}&=&\df14\lim_{\ep\to0}{\rm Tr}\int d\mu_3
\frac{1}{\psla{10}}\psla1\psla3\frac{1}{\psla{30}}\label{G3}\\
\ct{4}&=&\df14\lim_{\ep\to0}{\rm Tr}\int d\mu_2
\frac{1}{\psla1}\psla{12}\label{G4}
\end{eqnarray}
with $p_{i j}\equiv p_i-p_j$.

To proceed, we use the following properties of the measure\Eq{mu}:
\begin{eqnarray}
\int d\mu_1&=&-\frac{1}{\ep}\label{one}\\
\lim_{\ep\to0}\int d\mu_n&=&{2n \choose n}\zeta_{2n-1}\label{many}\\
\int d\mu_2\frac{p_0\cdot p_1}{p_1^2}&=&\frac{1+2\ep}{2}\int d\mu_2
\label{dm2a}\\
\int d\mu_2\frac{p_1\cdot p_2}{p_1^2}&=&\frac{1+\ep}{2}\int d\mu_2
\label{dm2b}
\end{eqnarray}
with\Eq{one} resulting from the choice of normalization in\Eq{mu},
and\Eqq{dm2a}{dm2b} from integration by parts.
The $n$-loop result\Eq{many},
with $n>1$, was proved in~\cite{LLP}, by
analysis of the wheel diagram with $n+1$ spokes in $D$ dimensions.
It generates the transcendentals associated~\cite{DK1,DK2}
with the $(2n-1,2)$ torus knots~\cite{VJ}, via crossed ladder
diagrams~\cite{DK2,UNI} that are obtained from wheel diagrams by
conformal transformation~\cite{LAD}. A purely four-dimensional derivation
of\Eq{many} was given in~\cite{LAD,5LB}, using Chebyshev-polynomial
techniques~\cite{GPX}.

These results lead to
immediate evaluation of\Eqq{G1}{G4}, for which\Eqqq{many}{dm2a}{dm2b}
give $\ct{1}=0$ and $\ct{4}=3\zeta_3$.
The four-term relation\Eq{4tr}
thus requires $\langle G_3-G_2 \rangle$ to evaluate to $3\zeta_3$,
which is a strong prediction for the three-loop two-point functions
of\Eqq{G2}{G3}, unexpected prior to~\cite{4TR}. We shall show that
each term evaluates to a multiple of the trefoil-knot transcendental,
$\zeta_3=\sum_{n>0}1/n^3$, and that the four-term relation is indeed
satisfied.

To complete the {\em experimentum crucis}, we use integration by
parts~\cite{CT} on the central triangles of $\ct{2}$ and $\ct{3}$
in Fig.~3. Each term so generated lacks a fermion propagator.
Subintegration then reduces the integrals to combinations of terms of
two-loop form, each with a propagator raised to a non-integer power.
This method is intrinsically $D$-dimensional; at $\ep=0$ separate
contributions diverge. Performing the subintegrations and relabelling momenta,
we obtain
\begin{equation}
\int d\mu_3\frac{1}{\psla{10}}\psla1\psla{k}\frac{1}{\psla{30}}
=\int d\mu_2\frac{1}{\psla{10}}H_k\frac{1}{\psla{20}}\label{ibp}
\end{equation}
for $k=2,3$, with
\begin{eqnarray}
(D-3)H_2&=&\frac{\psla1(\psla1+\psla2)(E_{10}-E_{12})}{\ep}+
(\psla0\psla1+\psla1\psla2)E_{10}\label{h2}\\
(D-4)H_3&=&\frac{2\psla1\psla2(E_{10}-E_{12})}{\ep}+2\psla0\psla2 E_{10}
\label{h3}
\end{eqnarray}
and $E_{i j}\equiv(p_0^2/p_{i j}^2)^\ep$.
Evaluation of $\ct{3}$, from $H_3$, thus requires one to
expand two-loop integrals to ${\rm O}(\ep^2)$. However,
we found that this did not generate $\zeta_5$, whose absence
is required by the four-term relation, since no other term may generate it.
Knot theory~\cite{DK1,DK2} alone is insufficient to show the absence
of $\zeta_5$, since the momentum flow in $\ct{2,3}$ is identical to that
in the four-loop zig-zag diagram~\cite{BKP} for renormalization of $\phi^4$
theory, which yields~\cite{CT} $20\zeta_5$, corresponding~\cite{DK1} to
the (5,2) torus knot~\cite{VJ}.

It remains to perform the two-loop integrals\Eq{ibp} and take the limit
$\ep\to0$ in\Eqq{G2}{G3}. This is easily accomplished with the aid
of~\cite{BGK}, where we showed how to reduce two-loop two-point integrals
to Saalsch\"{u}tzian ${}_3F_2$ series, when there are two adjacent propagators
with integer exponents. Such series were also encountered in~\cite{SJH,AVK}
and are exploited in~\cite{BGK,OSC,LNF,IMU3}.
Their $\ep$\/-expansions are easily developed~\cite{BGK}, using
identities systematized in~\cite{OSC}. Only at the level
relevant to six-loop renormalization~\cite{BKP}
does one first~\cite{1440,Z6} encounter a transcendental
that is an irreducible multiple zeta value~\cite{DZ}, of the type
studied by~\cite{EUL,BGK,BK15,IMU3}, in the context of field theory
and knot theory, and by~\cite{BBG,BG,BBB,MEH,PETA,BBBR}, in the context of
number theory. Up to five loops, all counterterms
are believed to be reducible to $\{\zeta_n\mid3\le n\le7\}$, though
an algorithm for achieving this reduction is established only up
to four loops~\cite{CT}.

Having already exploited integration
by parts, in\Eq{ibp}, before taking traces, we found it
more economical to use the hypergeometric recurrence relations of~\cite{BGK},
instead of the full machinery of~\cite{CT,CPC}.
To check our results, we used the {\sc Reduce}~\cite{RED} program
{\sc Slicer}~\cite{BKT}, which implements~\cite{CT}
by slicing four-loop  bubble diagrams, and hence avoids the
proliferation of three-loop two-point cases
that are handled in separate programs by {\sc Mincer}~\cite{CPC}.
Each method yields $\ct{2}=3\zeta_3$ and $\ct{3}=6\zeta_3$.

We hence verify the four-term relation of~\cite{4TR}, in its sole
non-trivial appearance below five loops, where the four-loop
diagrams of Fig.~3 give
\begin{equation}
\langle G_1-G_2+G_3-G_4\rangle=0-3\zeta_3+6\zeta_3-3\zeta_3=0\label{4lp}
\end{equation}
demonstrating cancellation of the trefoil knot in a manner that could
scarcely have been anticipated before the analysis of~\cite{4TR}.

\subsection*{3. Vector couplings and vector propagators}

In~\cite{4TR} the derivation of\Eq{4tr} made an explicit stipulation
that propagators adjacent to the chord have no tensor structure.
To investigate whether this restriction is indeed necessary,
we consider the case that $\times$ in Fig.~3 represents a $\gamma_\mu$
coupling to an external vector boson. This modifies the four terms as follows
\begin{eqnarray}
\ctt{1}&=&\df{1}{16}\lim_{\ep\to0}{\rm Tr}\int d\mu_2\gamma^\mu
\frac{1}{\psla1}\psla{12}\gamma_\mu\frac{1}{\psla{12}}\psla{02}
=\frac{3\zeta_3-1}{2}\label{Gt1}\\
\ctt{2}&=&\df{1}{16}\lim_{\ep\to0}{\rm Tr}\int d\mu_3\gamma^\mu
\frac{1}{\psla{10}}\psla1\psla2\psla3\gamma_\mu\frac{1}{\psla3}
\frac{1}{\psla{30}}=\frac{3\zeta_3+1}{2}\label{Gt2}\\
\ctt{3}&=&\df{1}{16}\lim_{\ep\to0}{\rm Tr}\int d\mu_3\gamma^\mu
\frac{1}{\psla{10}}\psla1\psla2\gamma_\mu\frac{1}{\psla2}
\psla3\frac{1}{\psla{30}}=-3\zeta_3\label{Gt3}\\
\ctt{4}&=&\df{1}{16}\lim_{\ep\to0}{\rm Tr}\int d\mu_2\gamma^\mu
\frac{1}{\psla1}\psla{12}\psla{02}\gamma_\mu\frac{1}{\psla{02}}
=-\df32\zeta_3\label{Gt4}
\end{eqnarray}
which, as allowed by the provisos,
fail to satisfy a four-term relation.

Similarly, we find that there is no four-term relation when
the chord is a vector boson, with any rational choice of
the gauge parameter $a$ in its propagator $g_{\mu\nu}/k^2
+(a-1)k_\mu k_\nu/k^4$.

\subsection*{4. Indications of richer structure at five loops}

There is one class of five-loop subdivergence-free counterterms
that may be obtained~\cite{LLP} from integration by parts: those whose
momentum flow is that of the wheel with five spokes.
Consider a putative four-term relation, generated by Fig.~4.

\begin{center}{\bf Fig.~4}\quad
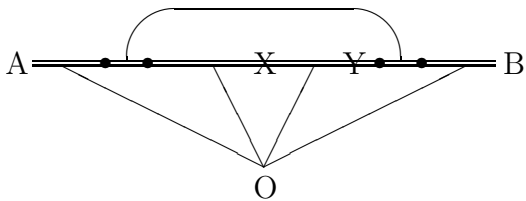
To generate four terms, at five loops, connect O to each blob, in turn.
\end{center}\begin{center}
\begin{picture}(400,200)(-200,-50)\put(0,100){\oval(260,100)[t]}
\put(-220,100){\line(1,0){440}}\put(-220,96){\line(1,0){440}}
\put(0,0){\line(-1,2){48}}\put(0,0){\line(1,2){48}}
\put(0,0){\line(2,1){192}}\put(0,0){\line(-2,1){192}}
\put(-150,98){\blob}\put(-110,98){\blob}\put(110,98){\blob}\put(150,98){\blob}
\put(-245,88){A}\put(227,88){B}\put(-10,-30){O}
\put(-10,88){X}\put(75,88){Y}\end{picture}
\end{center}

Each term is a radiative correction to a $\overline{\psi}\phi^2\psi$
coupling, induced by Yukawa couplings and a non-renormalizable $\phi^5$
interaction, thereby violating proviso~(v) of Section~2. After
systematic implementation of integration by parts for five-spoke
wheels, via recurrence relations on 15 exponents of Lorentz scalars,
we found that the counterterms
\begin{eqnarray}
\ctb{1}&=&\df14\lim_{\ep\to0}{\rm Tr}\int d\mu_3
\psla{1}\frac{1}{\psla{30}}=-2\zeta_3\label{G1b}\\
\ctb{2}&=&\df14\lim_{\ep\to0}{\rm Tr}\int d\mu_4
\frac{1}{\psla{10}}\psla1\psla2\frac{1}{\psla{40}}=4\zeta_3\label{G2b}\\
\ctb{3}&=&\df14\lim_{\ep\to0}{\rm Tr}\int d\mu_4
\frac{1}{\psla{10}}\psla1\psla4\frac{1}{\psla{40}}=20\zeta_5\label{G3b}\\
\ctb{4}&=&\df14\lim_{\ep\to0}{\rm Tr}\int d\mu_3
\frac{1}{\psla{10}}\psla1=10\zeta_5\label{G4b}
\end{eqnarray}
fail to satisfy a four-term relation. This failure (discovered by DJB)
was the origin of proviso~(v) in~\cite{4TR} and indicates how closely
the pure four-term relation is associated with {\em renormalizable\/}
field theory.

Remarkably, a four-term relation {\em is\/} obtained, if one moves the
external vertex Y, on the $p_4$ line of $\ctb{2}$, to the $p_3$ line
where X resides, giving
\begin{equation}
\langle\overline{G}_2^{\;*}\rangle=\df14\lim_{\ep\to0}{\rm Tr}\int d\mu_4
\frac{1}{\psla{10}}\psla1\psla2\frac{1}{\psla3}\psla4\frac{1}{\psla{40}}
=10\zeta_5-2\zeta_3\label{G2s}
\end{equation}
and hence non-trivial five-loop cancellation
\begin{equation}
\langle\overline{G}_1-\overline{G}_2^{\;*}+\overline{G}_3-\overline{G}_4
\rangle=-2\zeta_3-(10\zeta_5-2\zeta_3)+20\zeta_5-10\zeta_5=0\label{5lp}
\end{equation}
of both the $(3,2)$ and $(5,2)$ torus knots.
Efforts are in hand to derive the modified four-term
relation\Eq{5lp} from mixing of $\overline{\psi}\phi^2\psi$
radiative corrections with $(\overline{\psi}\psi)^2$ corrections that are
indistinguishable from the former, after nullification. For the present,
we adduce it as an indication of richer structure that
may be deducible from the extension of~\cite{4TR} to cases in which
the provisos are relaxed.

Finally, we remark on a specific {\em three\/}-term relation,
derived in~\cite{4TR}.
It is possible that such relations, called STU relations in the theory
of chord diagrams~\cite{BN}, impose even stronger constraints upon
the structure of field-theory counterterms.
Here we give a single intriguing example.
It involves the five-loop counterterms $\ctb{3,4}$, above,
which are related,
via $\langle\overline{G}_3-\overline{G}_4\rangle=\langle I_{\rm sub}\rangle$,
to a four-loop counterterm, $\langle I_{\rm sub}\rangle$,
that occurs because of a subdivergence in the contour integrals
that were devised in~\cite{4TR} to derive the four-term relation\Eq{4tr}.
Note that the counterterm $\ctb{4}=10\zeta_5$ is {\em also\/} reducible
to a four-loop diagram, since it contains a trivial convergent one-loop
subintegration.
Moreover, the identity $\ctb{4}\equiv\langle I_{\rm sub}\rangle$ is
obtainable purely at the diagrammatic level, without need of four-loop
integration. Hence the three-term relation of~\cite{4TR} tells us that
\begin{equation}
\ctb{3}=\ctb{4}+\ctb{4}\label{STU}
\end{equation}
which is {\em indeed\/} confirmed by the
highly non-trivial calculations\Eqq{G3b}{G4b}.
We are still recovering from our surprise at this successful
prediction of the five-loop counterterm\Eq{G3b}.
Before the advent of~\cite{DK1,DK2}, it might have been
expected to involve
$\zeta_3$, $\zeta_5$ and $\zeta_7$, in any rational combination.

We expect that the source of the findings above as well as the
restrictions summarized in the provisos can be ultimately
explained by the presence of a modified STU relation.
We expect this relation to connect the difference
between two three-point ccouplings to a four-point coupling.
We will report progress along these lines elsewhere \cite{dkfin}.

\subsection*{5. Conclusions}

We have used the methods of~\cite{CT,BGK} to verify the
sole predicted~\cite{4TR} non-trivial four-term relation between
subdivergence-free four-dimensional counterterms with less than five loops.
Analytical tools~\cite{ai92} do not yet exist~\cite{ai93} to
investigate pure four-term relations in four-dimensional
renormalizable theories at five loops and beyond, where trivalent
couplings frustrate the progress that we achieved to seven loops~\cite{BKP}
in pure $\phi^4$ theory. Nor can the all-order methods
of~\cite{BGK} be turned to account, at present,
since these derive from large-$N$ methods, where
subdivergences are of the essence.
The situation is somewhat similar to that in quenched
QED, where the cancellation of transcendentals
that is predicted to all orders by knot theory~\cite{BDK}
has been confirmed at four loops~\cite{4LQ}, with little immediate
prospect of progressing to five loops.

In a forthcoming book~\cite{DK},
the relation to three-dimensional Chern-Simons theory will be discussed.
{}From the calculational point of view, this topological theory (with
apparently no possibility of observable particles) appears to have little to
offer in terms of multi-loop perturbative results, which is ironical
in view of the fact that much discussion of four-term relations
has so far been grounded in its chord diagrams~\cite{BN,JKTT}.
We note that the rational three-dimensional two-loop beta-function
in~\cite{Tan} is in full accord with the expectations of~\cite{DK1,DK2}.
A promising avenue of multi-loop inquiry concerns the cosmological constant
generated by Yukawa and $\phi^4$ couplings, whose four-loop analysis
proved tractable in three dimensions~\cite{3D},
where it is a logarithmically divergent quantity, because of the
super-renormalizability of the theory.

If one is prepared to progress to five loops, with purely
trivalent couplings, then renormalization of $\phi^3$ theory in six
dimensions~\cite{DEM} is the cleanest case to study, being free of any
tensorial complication, and extremely benign in the infrared.
Here, one suspects that the first tests of four-term relations will
be made by approximation methods, rather than analytically.

More generally, we have an intuition that worldline
techniques~\cite{CS} may illuminate cancellation of
knot-transcendentals between counterterms, in both the quenched-QED
analysis of~\cite{BDK} and also the four-term analysis of~\cite{4TR},
since each is concerned with subdivergence-free combinations of
chord diagrams. The results in \cite{hopf} indicate that the
incorporation of diagrams with subdivergences is not out of reach.

In conclusion, it is gratifying that the
prediction of~\cite{4TR} is borne out in the only non-trivial
test that we have been able to devise, without exorbitant
labour. It is, however, still frustrating to lack further
definitive case law, beyond\Eq{4lp} and
the expected failures of Sections~3 and~4.
The modified four-term relation\Eq{5lp}
and the remarkable STU-type relation\Eq{STU}
indicate that an even richer structure of counterterms awaits discovery.
We hope that colleagues will exercise ingenuity to progress the issue,
either by developing~\cite{CS} calculational techniques that
may prove more powerful than
the four-loop methods of~\cite{CT}, and less restricted than
the $n$-loop methods of~\cite{LLP},
or by analyzing how the four-term relation is modified by subdivergences,
by vector couplings, by vector propagators,
and by dimensionful coupling constants.
Some progress along these lines will be reported elsewhere \cite{dkfin}.

\noindent{\bf Acknowledgements}\quad
We are enormously grateful to Bob Delbourgo,
for generous advice and hospitality during
our stay at the University of Tasmania in July and August, thanks to
which the present work and seven other major
projects~\cite{4TR,BGK,BK15,DK3,IMU3,PETA,DK} flourished.
We thank our field-theory~\cite{BGK,BDK,IMU3} co-authors, Bob Delbourgo,
John Gracey and Tolya Kotikov, and our number-theory~\cite{BBB,BBBR}
co-authors, Jon Borwein, David Bradley and Roland Girgensohn,
for their encouragement.
The interest of Tim Baker, Peter Jarvis, David McAnally and
Ioannis Tsohantjis stimulated us to compute the four-loop result\Eq{4lp}
in Hobart, with assistance from Neville Jones.
Assistance from Chris Stoddart enabled computation of the
five-loop results\Eqq{5lp}{STU} on an AlphaCluster at the Open University.
The support of our partners, Margaret and Susan, was inestimable.

\raggedright

\end{document}